\documentclass[letter]{aa} 
\usepackage{deluxetable}
\usepackage{times,epsfig} 
\usepackage{mathrsfs}  
\usepackage{natbib}
\usepackage{amssymb}
\usepackage{longtable}
\bibpunct{(}{)}{;}{a}{}{,}
\usepackage{caption}
\usepackage[varg]{txfonts}
\usepackage[section]{placeins}
\usepackage[english]{babel}
\usepackage{url}
\usepackage{hyperref}
\usepackage{arydshln}
\usepackage{stfloats}
\usepackage{color}
\usepackage{amsmath}

\def\phs{\phantom{$-$}}

\authorrunning{Polshaw et al.}
\titlerunning{SN~2014bc in NGC~4258}

\begin{document}

\title{A supernova distance to the anchor galaxy NGC~4258}

\author{
J. Polshaw\inst{1}{\thanks{E-mail: jpolshaw01@qub.ac.uk}}
\and R. Kotak\inst{1}
\and K.C. Chambers\inst{2}
\and S.J. Smartt\inst{1}
\and S. Taubenberger\inst{3,}\inst{4}
\and M. Kromer\inst{5}
\and E.E.E. Gall\inst{1}
\and W. Hillebrandt\inst{4}
\and M. Huber\inst{2}
\and K.W. Smith\inst{1}
\and R.J. Wainscoat\inst{2}
}

\institute{Astrophysics Research Centre, School of Mathematics and Physics, Queen’s University Belfast, Belfast BT7 1NN, UK
\and 
Institute for Astronomy, University of Hawaii, 2680 Woodlawn Drive, Honolulu, HI 96822, USA
\and
European Southern Observatory, Karl-Schwarzschild-Str. 2, D-85748 Garching bei München, Germany
\and 
Max-Planck-Institut fur Astrophysik, Karl-Schwarzschild-Str. 1, D-85748 Garching bei M{\"u}nchen, Germany
\and 
The Oskar Klein Centre \& Department of Astronomy, Stockholm University, AlbaNova, SE-106 91 Stockholm, Sweden
}

\date{Received ... / Accepted ...}
\abstract{
The fortuitous occurrence of a type II-Plateau (IIP) supernova, SN~2014bc, in a galaxy for which distance estimates from a number of primary distance indicators are available provides a means with which to cross-calibrate the standardised candle method (SCM) for type IIP SNe. By applying calibrations from the literature we find distance estimates in line with the most precise measurement to NGC~4258 based on the Keplerian motion of masers (7.6$\pm$0.23\,Mpc), albeit with significant scatter. 
We provide an alternative local SCM calibration by only considering type IIP SNe that have occurred in galaxies for which a Cepheid distance estimate is available. We find a considerable reduction in scatter ($\sigma_I = 0.16$\, mag.), but note that the current sample size is limited. Applying this calibration, we estimate a distance to NGC~4258 of $7.08\pm0.86$ Mpc.  
}


\maketitle

\section{Introduction}

\label{sec:intro}
Although the use of thermonuclear (type Ia) supernovae (SNe) as standardisable 
candles is now firmly embedded within the fabric of mainstream cosmology, consideration
of other types of SNe as either relative or absolute distance indicators is still in its
infancy. The reasons for this are manifold, ranging from overall faintness at epochs of interest
\citep[e.g. type IIP SNe;][]{Hamuy_2002,Nugent_2006}, to a current lack of comparable precision 
when compared to SNe Ia \citep[e.g. GRB-SNe;][]{CJ:14,Li_2014} or superluminous SNe \citep{IS:14}.
Of the non-type Ia SNe, arguably the most promising avenue is afforded by the type IIP 
SNe, as their relative faintness is amply compensated by their higher frequency per unit volume, 
while the tight correlation between ejecta velocity and plateau brightness 
\citep[`standardised candle method', hereafter SCM;][]{Hamuy_2002} rests on physically-motivated, and 
well-understood grounds.

The local distance scale hinges primarily on the calibration of the Cepheid period-luminosity ($P-L$) relation to the Large Magellanic Cloud (LMC) distance. 
Although no effort has been spared in attempting to quantify the systematic effects that affect the $P-L$ relation, \citep[e.g.][]{freedman:01,fausnaugh:15}, the suitability of LMC as the first rung on the cosmological distance ladder has itself been called into question \citep[e.g.][]{macri:06,Riess_2011}. 
NGC~4258 is a natural choice for anchoring the Cepheid distance scale as a geometric maser distance with an uncertainty of only 3\% has been measured \citep[7.6$\pm$0.23\,Mpc;][]{Humphreys_2013}.

In this {\it Letter}, we use SN~2014bc to obtain the distance to NGC~4258, the only galaxy for which both maser and Cepheid distance measurements are available.

\section{Observations}

\begin{figure}[t]
\centering
\resizebox{\hsize}{!}{\includegraphics[width=17cm]{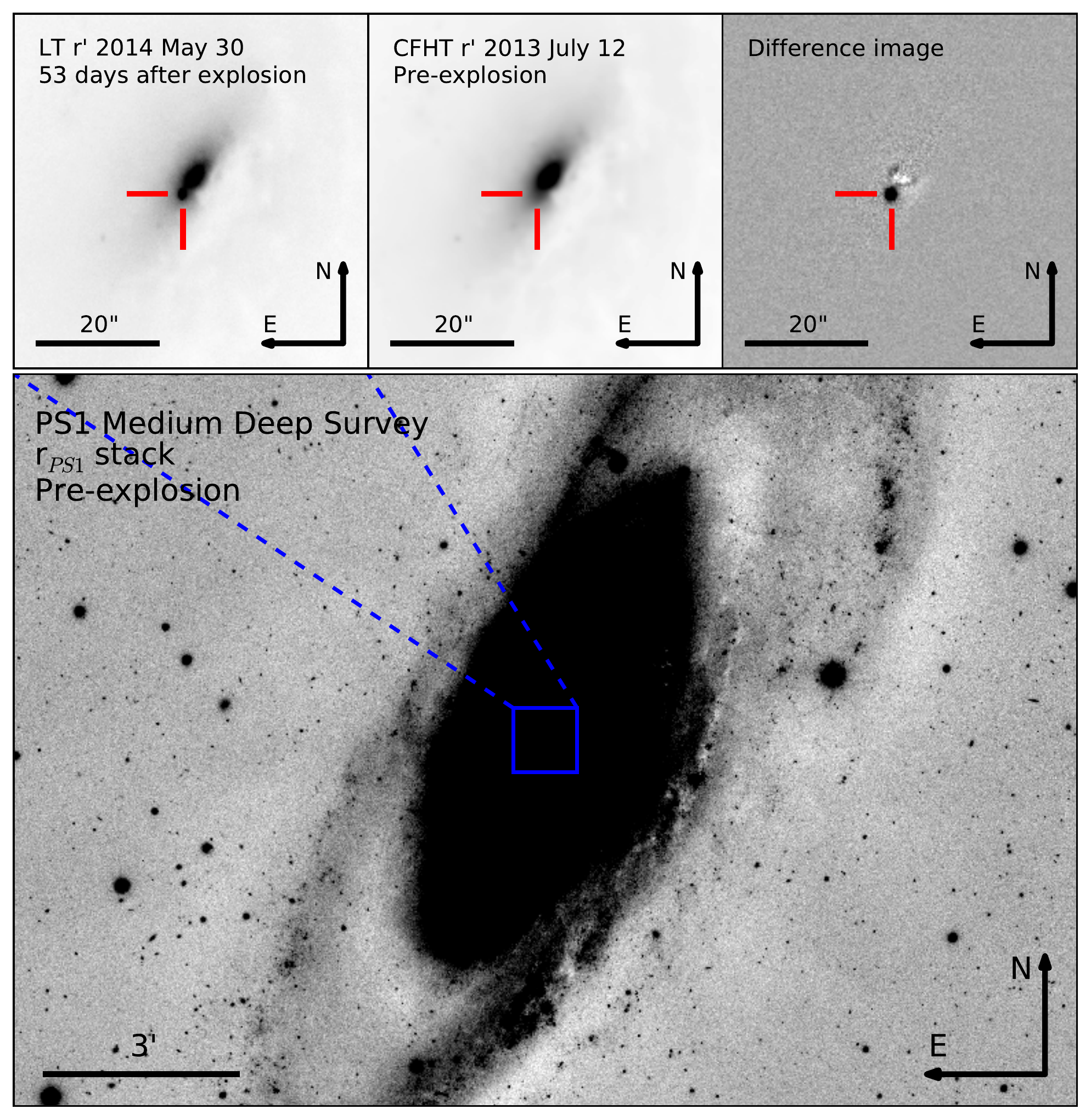}}
\caption{
\textbf{Top:} $r'$-band image of SN~2014bc (left), pre-explosion image (middle), and the difference between the two (right). SN~2014bc is located at $\alpha_\mathrm{J2000}$=12:18:57.71, $\delta_\mathrm{J2000}$=+47:18:11.3, projected $3\farcs67$ ($\simeq 135$ pc, at 7.6\,Mpc) from the nucleus of NGC~4258. The position of the SN is marked by red dashes.
\textbf{Bottom:} Stacked $r_{\mathrm{PS1}}$ image of NGC~4258. All images included in the stack were taken prior to 2014 Apr. 08. The $1' \times 1'$  blue box delineates the region shown in the upper panels.
\label{fig1}}
\end{figure}

SN~2014bc in NGC~4258 was discovered by the Panoramic Survey Telescope and Rapid Response System 1 (PS1; \citealt{Kaiser_2010}) on 2014 May 19.25 \citep{ATel6156}. However, upon further inspection, the first PS1 detection was in fact on 2014 Apr 11.40, when NGC~4258 was serendipitously observed during the PS1 NEO survey.
The Lick Observatory Supernova Search (LOSS) subsequently reported an earlier detection on 2014 Apr 9.35 \citep{ATel6159}, although they do not give a precise magnitude due to the proximity of the SN to the bright nucleus of NGC~4258 (see Fig. \ref{fig1}). Nevertheless, LOSS did report a non-detection on 2014 Apr 6.35 at $R > +18.0$ mag., allowing us to constrain the explosion epoch to 2014 Apr 7.9 $\pm$ 1.5 (MJD = 56754.9 $\pm$ 1.5), which we will hereafter refer to as day $0$. An optical spectrum  obtained 2\,d after discovery (+43.9\,d)  revealed the object to be a type-II SN \citep{Ochner_2014}. 

\begin{figure}[t]
\centering
\resizebox{\hsize}{!}{\includegraphics[width=17cm]{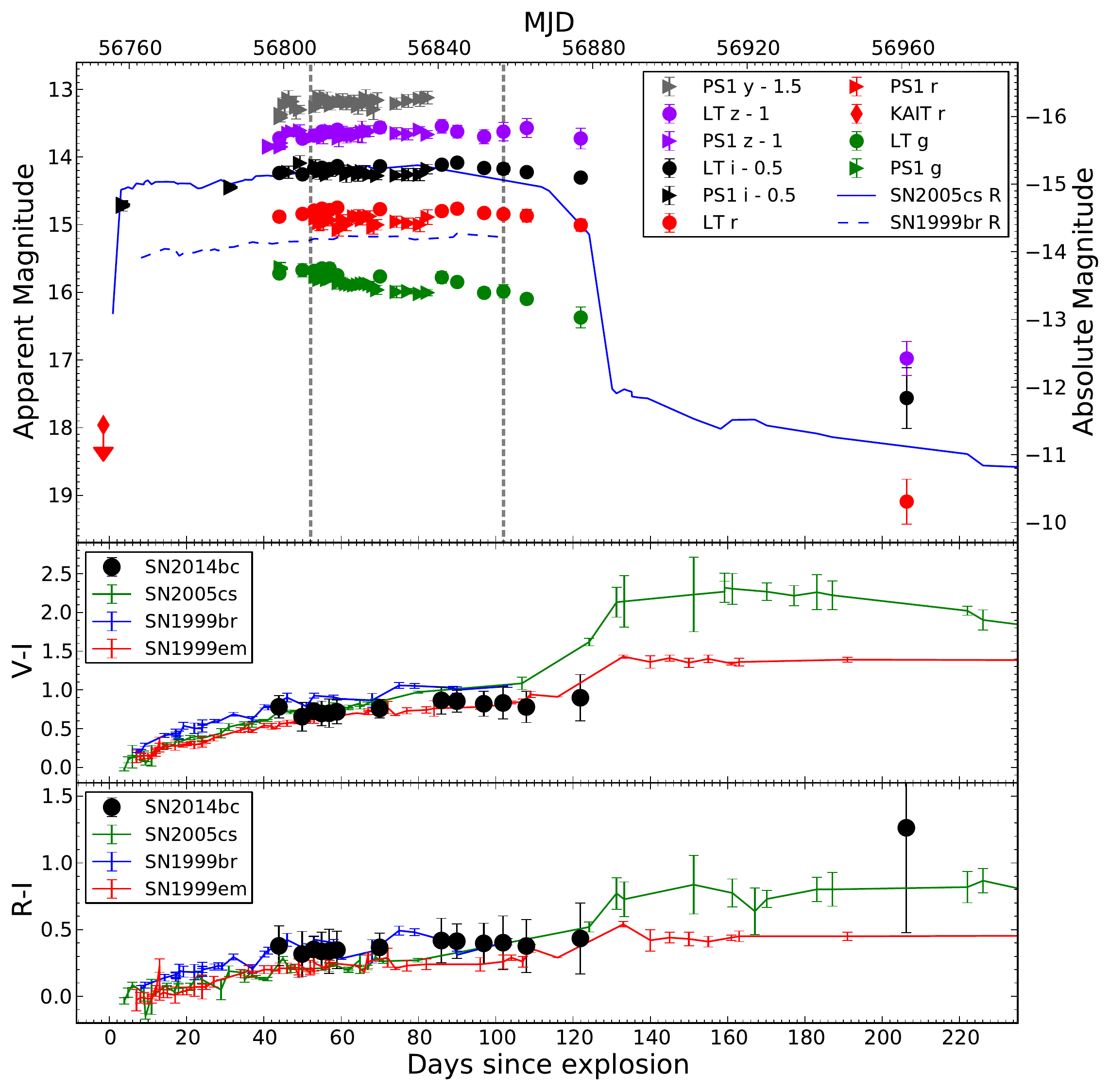}}
\caption{
\textbf{Top:} Light curves of SN~2014bc, corrected for Galactic reddening. The absolute magnitude scale was determined using $\mu = 29.4$\,mag.
The $R$-band light curves of the type IIP SNe 1999br and 2005cs 
are  shown according to the absolute magnitude scale. Downward pointing arrows indicate limiting magnitudes. The vertical dashed lines indicate the spectroscopic epochs of SN~2014bc. Small differences between the SDSS and PS1 photometric systems are evident \citep{Tonry_2012b}.
\textbf{Middle and bottom:} Colours of SNe 2014bc compared to SNe 1999br, 2005cs, and 1999em. 
\label{fig2}}
\end{figure}

We obtained followup optical imaging of the SN using a combination of the 2m Liverpool Telescope (LT) in the $g'r'i'z'$ filters, and PS1 in the $g_{\mathrm{PS1}}\,r_{\mathrm{PS1}}\,i_{\mathrm{PS1}}\,z_{\mathrm{PS1}}\,y_{\mathrm{PS1}}$ filters. The images were reduced in a standard fashion by the LT and  PS1 Image Processing Pipelines (IPP; \citealt{Magnier_2013}), respectively. The $griz$ zeropoints and colour terms were measured using a sequence of stars in the SDSS DR9 catalogue, while the PS1 zeropoints were computed by the IPP and taken from the images \citep{Schlafly_2012,Magnier_2013}.
As SN~2014bc is located just $3\farcs67$ from  the bright nucleus of NGC~4258 (Fig. \ref{fig1}), direct
measurements of the SN brightness were rendered difficult. 
We therefore subtracted archival pre-explosion images from each followup optical image using {\sc hotpants}\footnote{http://www.astro.washington.edu/users/becker/hotpants.html}. For the LT $g,r,$ and $i$ data we used images taken by the Canada France Hawaii Telescope (CFHT)
taken on 2005 Apr 06, 2013 Jul 12 and 2009 Feb 26 respectively. For the LT $z$-band we used the SDSS image of the field (taken on 2002 Dec 13); for the PS1 data we used the Medium-Deep Survey stacks \citep{Tonry_2012}. An example of template subtraction is shown in the upper panels of Fig. \ref{fig1}. The magnitude of the SN in the difference image was measured via PSF-fitting using the SNOoPY\footnote{http://sngroup.oapd.inaf.it/snoopy} package within {\sc iraf}.
The photometry of SN~2014bc is given in Table \ref{table:photmetry}, and the light curves are shown in the top panel of Fig. \ref{fig2}.

\begin{figure*}[!t]
\centering
\resizebox{\hsize}{!}{\includegraphics[width=\textwidth]{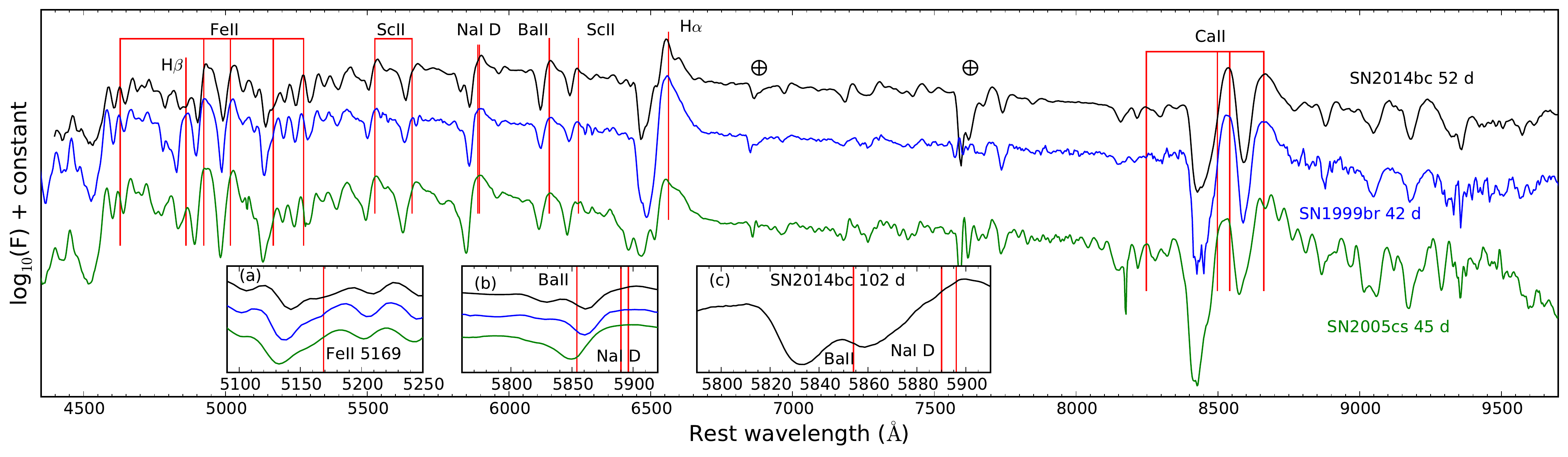}}
\caption{Optical spectra of SNe 2014bc, 1999br, and 2005cs, all at approximately 50\,d after explosion. The spectra have been corrected for reddening and redshift. 
The $\oplus$ symbol indicates significant telluric absorption. 
Insets (a) and (b) show zoom-ins of the \ion{Fe}{ii} $\lambda$5169 and \ion{Na}{i} D regions respectively. Inset (c) shows a zoom-in of the higher resolution 102\,d spectrum of SN~2014bc. The rest wavelengths of \ion{Fe}{ii} $\lambda$5169, \ion{Ba}{ii} $\lambda$5854 and the \ion{Na}{i} $\lambda\lambda$5890, 5896 doublet are marked by vertical red lines in the inset panels.
\label{fig3}} 
\end{figure*}

We obtained an optical spectrum of the SN at +52\,d with the Gran Telescopio Canarias + OSIRIS. The R300B grism was used with a slit width of $1\farcs0$, and a total exposure time of 525\,s over four separate exposures, which led to a wavelength range of 4400 -- 10000\,$\AA$ at a resolution of 17.4\,$\AA$ as measured from the FWHM of the [\ion{O}{i}] $\lambda$5577$\AA$ sky line (average of all exposures). The spectrum was reduced using standard techniques, with relative flux calibration performed using the spectrophotometric standard star GD 153, and an absolute scaling achieved from coeval photometry.
This spectrum is presented in Fig. \ref{fig3}. We also obtained a spectrum on day 102 with the R1000B grism and a $0\farcs8$ slit that yielded a resolution of 9.4\,$\AA$.
Both spectra were taken with identical slit position-angles, and aligned so as to minimize contamination from the bright nucleus of NGC~4258.

\subsection{Reddening estimate and absolute magnitudes}
\label{sec:red}

The Galactic reddening in the line of sight to SN~2014bc is $E(B-V) = 0.0141$ mag \citep{Schlafly_2011}. 
In order to estimate the host galaxy reddening, we consider the colour evolution, spectral shape, and search for narrow \ion{Na}{i} D absorption.
For type IIP SNe, it is reasonable to assume that the photospheric temperature should not drop below the hydrogen recombination temperature during the plateau phase. Although the actual composition will play a role, to zeroth order all type IIP SNe should evolve through the same colour at the end of the plateau \citep{eastman:96,Hamuy_2003a}. In the middle and lower panels of Fig. \ref{fig2}, we compare the colour evolution of SN~2014bc to other II-P SNe. We find that the SN~2014bc colours -- only corrected for Galactic reddening -- closely track those of SNe 1999br, 1999em, and 2005cs, which were inferred to have a total $E(B-V)$ of 0.024, 0.1, and 0.05 mag., respectively \citep{Pastorello_2004,elmhamdi:03,Pastorello_2009}. Our photometric coverage of SN~2014bc does not extend to the transition off the plateau phase given that it was reaching conjunction with the sun at that epoch. Nevertheless, if we assume that our set of points at 122\,d are at, or close to the end of the plateau, then the colours are entirely consistent with other, virtually unreddened type II-P SNe, which have observations that do continue beyond the end of the recombination phase. 

Another commonly-used technique to estimate the host galaxy reddening in SNe is to measure the equivalent widths of the narrow  interstellar absorption lines due to \ion{Na}{i} D \citep[e.g.][]{Turatto_2003}. We do not detect narrow \ion{Na}{i} lines in any of our spectra at the redshift of NGC~4258 (see insets in Fig. \ref{fig3}), including our highest resolution spectrum (9.4\,\AA). 
This implies little to no reddening due to the interstellar medium of the host galaxy. 
\citet{Ochner_2014} report narrow \ion{Na}{i} lines in their admittedly low signal-to-noise and heavily contaminated classification spectrum taken only 8\,d prior to our first spectrum. Upon inspection of this publicly available spectrum via the Asiago Transient Classification Programme, we speculate that this might 
be attributable to differences in the slit positioning, seeing, and spatial resolution.

Examining all the evidence in hand, we note that the colour evolution of SN~2014bc during the plateau is completely consistent with other type IIP SNe, while the absolute magnitudes are intermediate between those of SNe~1999br and 2005cs, assuming negligible host reddening (Fig. \ref{fig2}). Furthermore, the spectral energy distribution of our 52\,d spectrum matches these two comparison SNe, without requiring any additional reddening correction (Fig. \ref{fig3}). These similarities, taken together with the absence of narrow \ion{Na}{i} D absorption lines point towards little host reddening in SN~2014bc, and in what follows, we treat it as negligible, and correct only for Galactic reddening. 
At first glance, this may be in stark contradiction to expectations, given the location of the SN. However, in the case of SN~2014bc, it is likely that its small galactocentric distance is merely a projection in front of the nucleus, rather than an indication that it is embedded in the galaxy core; other examples include SNe 2010cu and 2011hi, both of which occurred at projected distances of $\lesssim 400$\,pc from the centre of the dusty, luminous infrared galaxy IC~883 (100\,Mpc), but had low inferred host galaxy extinctions \citep{Kankare_2012}. 
Thus, using the maser distance modulus to NGC~4258, we find the absolute average plateau brightness to be $M_{r} = -14.48$\,mag and $M_{i} = -14.65$\,mag.

\section{SCM applied to SN~2014bc}

A tight, positive correlation between the (quasi-) bolometric luminosity and expansion velocity 
during the plateau phase of type IIP SNe allows them to be used as standardised candles
\citep{Hamuy_2002}. The mid-point of the plateau ($\sim$50\,d) is traditionally used as a convenient reference point.

In order to apply this method to SN~2014bc, we first converted the $gri$ magnitudes of SN~2014bc to $VRI$ using the transformations of \citet{Jester_2005}, and fit low order polynomials to the resulting light curves to estimate the magnitudes at 50\,d. The resulting magnitudes are $m_V=+15.32 \pm 0.10$ mag and $m_I=+14.60 \pm 0.12$ mag (uncorrected for reddening). As a check, we computed synthetic $m_{V}$ and $m_{I}$ magnitudes from the 52\,d spectrum using the {\sc sms} code (Synthetic  Magnitudes from Spectra) within the S3 package (Inserra et al. \textit{in prep.}), which we found to be  consistent within the uncertainties of the above transformations. The expansion velocities of type II SNe are generally measured using weak lines as these are better tracers of the photospheric velocity, with the \ion{Fe}{ii} $\lambda$5169 line typically being used. However, in SN~2014bc, this line does not form a clean P-Cygni profile, and appears to suffer from  blending (see inset in Fig. \ref{fig3}). We therefore used the weighted mean of the \ion{Fe}{ii} $\lambda$4629, $\lambda$4924, $\lambda$5018, and $\lambda$5276 lines instead. We normalised the spectrum by fitting a low order polynomial to the pseudo-continuum and subtracting it from the spectrum. The velocity of each line was then estimated by measuring the position of the minimum of the absorption component with respect to the rest wavelength, by fitting Gaussian profiles. The procedure was repeated several times across slightly different wavelength ranges. Each line was then weighted by the standard deviation, yielding a velocity of $v_{\ion{Fe}{ii}}=1460\pm100$ km s$^{-1}$.

Next, we applied a series of SCM calibrations derived from different samples of type IIP SNe. Each calibration is based on relations between the expansion velocity at 50\,d post-explosion, the coeval $I$-band magnitude ($m_I$), and a reddening term.
The resulting distances for SN~2014bc from the calibrations are given in Table \ref{table:distances}. 
The differences between the SCM and maser distance moduli are provided, along with the median distance of the sample of SNe used in each calibration. These $\Delta\mu$ values are related to the corrections to the zero points required to bring the calibrations into agreement agreement with the maser distance.
Although there is a spread in the values, all calibrations are reassuringly consistent with that inferred from at least one of the primary distance indicators. In particular, only the \citet{Nugent_2006} calibration which excludes high-redshift SNe, marked `2b' in Table \ref{table:distances}, is not consistent with the maser distance. 

In order to further investigate the scatter among the various calibrations, we applied the SCM to all type IIP SNe that occurred in galaxies for which Cepheid distances have been measured, and for which a spectrum taken approximately at the mid-plateau point was available (Table \ref{table:IIP_cepheid}). The differences between the SCM and Cepheid distance moduli are listed in Table \ref{table:IIP_ZPs}. We find that the \citet{Hamuy_2003} and \citet{Nugent_2006} calibrations provide distance estimates that are closest to the Cepheid distances; however, most of the differences across all calibrations are within the typical uncertainty of $\sim$0.3 mag. 
All SCM calibrations yield a distance estimate to SN~2004dj -- the nearest SN in our sample -- that is longer than the corresponding Cepheid distance to the host galaxy. We speculate that this may be related to the late discovery of SN~2004dj, coupled with the uncertainty in its explosion epoch \citep{Zhang_2006}.
A Hubble diagram of these objects, using the Cepheid measurements and \citet{Riess_2011} $H_{0}$ value to fix the distance scale (Fig. \ref{figA1}), shows a remarkably small scatter ($\sigma_I \sim 0.16$\,mag) after applying the following SCM correction:
\begin{multline}\label{eq:SCM_I}
\mathrm{I_{50}-A_{I}+5.665(\pm0.487)log_{10}(v_{50}/5000)} \\ \mathrm{=5log_{10}(H_{0}D)-2.045(\pm0.137)}
\end{multline}
This calibration provides a distance estimate to SN~2014bc of $7.08\pm0.86$\,Mpc, in agreement with the maser distance.

\begin{table}[!t]
\setlength{\tabcolsep}{4pt}
\caption{Distance estimates to NGC~4258 }
\label{table:distances} 
\centering
\begin{tabular}{lcccc}
\hline\hline 
Method 	&	D & $\mu$	&	$\Delta\mu$ 	&	Median$^\dag\mu$	\\
& (Mpc) & (mag) & (mag) & (mag) \\
\hline
SCM$^{1\mathrm{a}}$	&	6.37 (1.13)	&	29.02 (0.38)	&	$-0.38$	&	32.0	\\
SCM$^{1\mathrm{b}}$	&	7.07 (1.15)	&	29.25 (0.35)	&	$-0.15$	&	34.9		\\
SCM$^{2\mathrm{a}}$	&	6.78 (1.01)	&	29.16 (0.32)	&	$-0.24$		&	34.2 \\
SCM$^{2\mathrm{b}}$	&	5.39 (1.05)	&	28.66 (0.42)	&	$-0.74$	&	33.2	\\
SCM$^{3}$	&	7.48 (1.07)	&	29.37 (0.31)	&	$-0.03$		&	32.7	\\
SCM$^{4}$	&	8.77 (1.45)	&	29.71 (0.36)	&	\phs0.31	&	34.8 \\	
\hline 
Maser$^{5}$	&	7.60 (0.23)	&	29.40 (0.07)	&	\phs\ldots	&	\phs\ldots	\\
Cepheid$^{6}$	& 7.40 (1.16)	&	29.35 (0.34)	&	\phs\ldots &	\phs\ldots	\\
\hline
SCM$^{7}$ (here)	&	7.08 (0.86)	&	29.25 (0.26)	&	\phs0.15	&	29.7	\\
\hline
\end{tabular}
\tablefoot{
$H_{0}=73.8$ km s$^{-1}$ Mpc$^{-1}$ \citep{Riess_2011}. 
Uncertainties are given in parentheses. 
$\Delta\mu$ is the difference between each SCM distance modulus estimate and the maser distance modulus.
$^\dag$Median distance modulus of the SNe comprising each sample.
1a) \citet{Hamuy_2003} Eq. 3, whole sample ($\sigma=0.32$mag); 
1b) \citet{Hamuy_2003} Eq. 4, excluding SNe with $z<0.01$ which results in a lower scatter ($\sigma=0.29$mag); 
2a) \citet{Nugent_2006} Eq. 1, whole sample; 
2b) \citet{Nugent_2006} Eq. 1, excluding SNe with $z>0.05$; 
3) \citet{Poznanski_2009} Eq. 2; 
4) \citet{DAndrea_2010} Eq. 2; the \citet{Poznanski_2009} sample is essentially subsumed in its entirety
in the computation of this calibration.
5) \citet{Humphreys_2013}; 
6) \citet{Fiorentino_2013} provide the most recent Cepheid distance estimate to NGC~4258;
7) This work: Eq. \ref{eq:SCM_I}.
}
\end{table}

\section{Conclusions}

\begin{figure}[t]
\centering
\resizebox{\hsize}{!}{\includegraphics[width=\textwidth]{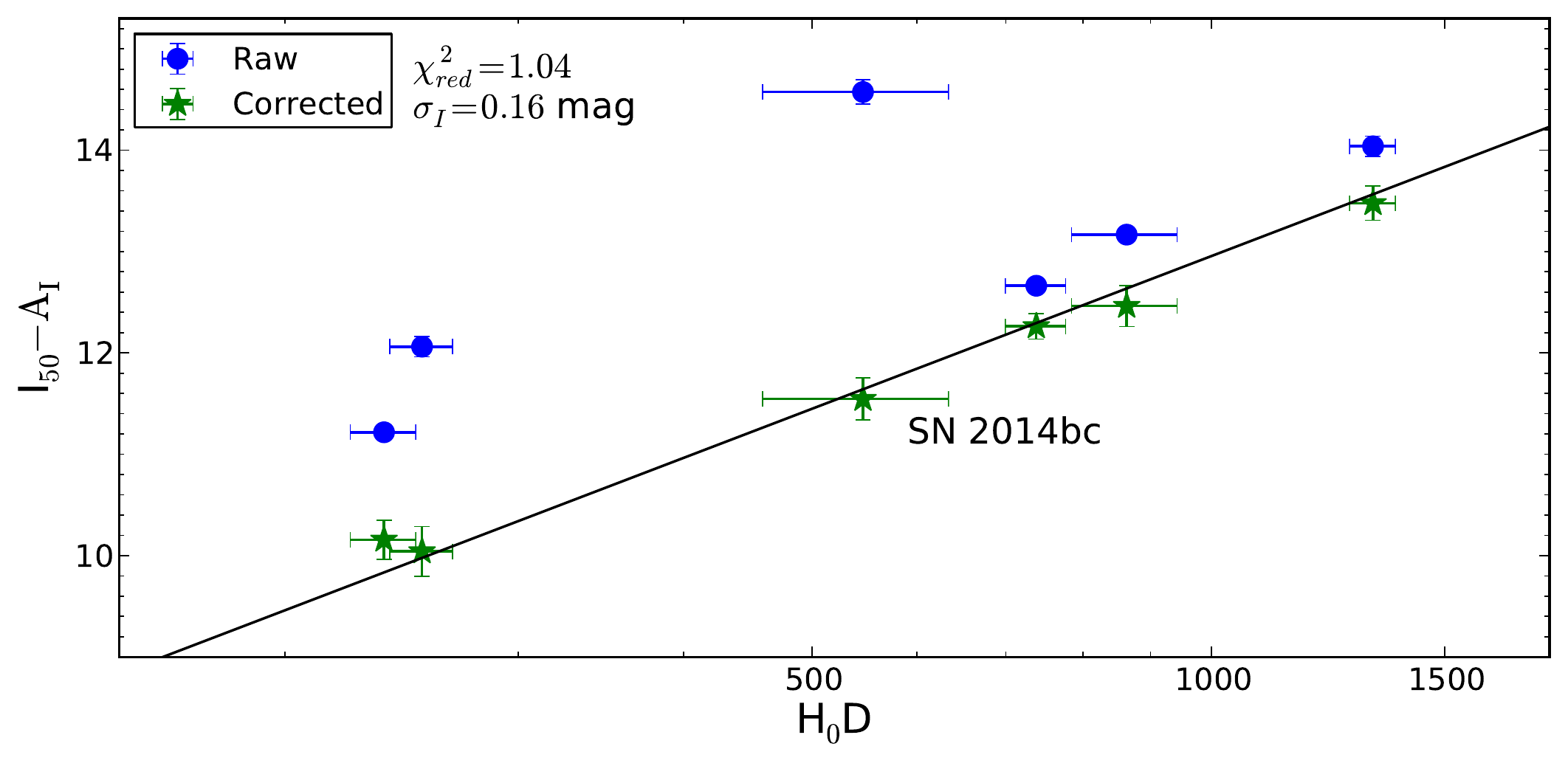}}
\caption{
Hubble diagram of SNe listed in Table \ref{table:IIP_cepheid}, using the Cepheid distances and $H_{0}$=73.8 km s$^{-1}$ Mpc$^{-1}$. The blue circles are corrected for reddening only, while the green stars are corrected for expansion velocities via least-squares fitting. 
\label{figA1}}
\end{figure}

Based on measurements of SN~2014bc, we find the SCM distance to the anchor galaxy NGC~4258 to be encouragingly consistent with the geometric maser distance, even though there is considerable scatter amongst the various calibrations. 
The current number of type IIP SNe occurring in galaxies with a previously measured Cepheid distance is surprisingly small. However, as for the SNe Ia, we expect the type IIP SN systematics to be better quantified in the coming years. 
Natural and immediate next steps, but ones that go beyond the scope of this work, would be to readjust the Cepheid-based distances to the galaxies listed in Table \ref{table:IIP_cepheid} using self-consistent $P-L$ calibrations and the revised LMC Cepheid distance modulus \citep{fausnaugh:15}.
Our expectation would be one of a knock-on further reduction in scatter for the various SCM calibrations when anchored to a sample akin to the one we present in Table \ref{table:IIP_ZPs} and Fig. \ref{figA1}.

\begin{acknowledgements}

We thank B. Leibundgut for useful comments on an earlier draft.
RK acknowledges support from STFC via ST/L000709/1.
SJS acknowledges funding from the ERC via Grant agreement n$^{\rm o}$ [291222].
Support by the Deutsche Forschungsgemeinschaft via the Transregional Collaborative Research Center TRR 33 
`The Dark Universe’ is acknowledged.
Based in part on observations carried out with the Gran Telescopio Canarias (GTC2007-12ESO), and the Liverpool Telescope installed in the Spanish Observatorio del Roque de los Muchachos of the Instituto de Astrofísica de Canarias, on the island of La Palma.
The LT is operated by Liverpool John Moores University with financial support from STFC.
Based in part on observations made with ESO Telescopes at the La Silla Paranal Observatory under programme ID 081.D-0128.
PS1 is supported by NASA Grants NNX12AR65G and NNX14AM74G from NEO Observation Program.
This paper includes observations obtained with MegaPrime/MegaCam, a joint project of CFHT and CEA/DAPNIA, at the Canada-France-Hawaii Telescope (CFHT), which is operated by the National Research Council (NRC) of Canada, the Institut National des Science de l’Univers of the Centre National de la Recherche Scientifique (CNRS) of France, and the University of Hawaii.

\end{acknowledgements}

\bibliographystyle{aa}
\bibliography{bibtex}

\appendix

\section{Tables}

\onecolumn

\begin{table}[t]
\caption{Photometry of SN~2014bc. Uncertainties are in parentheses. The epoch is given with respect to the estimated explosion date: MJD = 56754.9.}
\label{table:photmetry} 
 \centering
 \def\arraystretch{0.9}
 \begin{tabular}{lcccccccc}
 \hline\hline 
 Date	&	MJD	&	Epoch	&	$g$ / $g_{\mathrm{PS1}}$	&	$r$ / $r_{\mathrm{PS1}}$	&	$i$ / $i_{\mathrm{PS1}}$	&	$z$ / $z_{\mathrm{PS1}}$	&	$y_{\mathrm{PS1}}$	&	Telescope	\\
yyyy mm dd	&		&	(days)	&		&		&		&		&		&		\\
\hline
2014 04 11	&	56758.40	&	\phs\phs3.50	&	\ldots	&	\ldots	&	15.34 (0.07)	&	\ldots	&	\ldots	&	PS1	\\
2014 04 11	&	56758.41	&	\phs\phs3.51	&	\ldots	&	\ldots	&	15.30 (0.05)	&	\ldots	&	\ldots	&	PS1	\\
2014 04 11	&	56758.43	&	\phs\phs3.53	&	\ldots	&	\ldots	&	15.30 (0.03)	&	\ldots	&	\ldots	&	PS1	\\
2014 04 11	&	56758.44	&	\phs\phs3.54	&	\ldots	&	\ldots	&	15.31 (0.02)	&	\ldots	&	\ldots	&	PS1	\\
2014 05 09	&	56786.26	&	\phs31.36	&	\ldots	&	\ldots	&	15.07 (0.01)	&	\ldots	&	\ldots	&	PS1	\\
2014 05 09	&	56786.27	&	\phs31.37	&	\ldots	&	\ldots	&	15.04 (0.02)	&	\ldots	&	\ldots	&	PS1	\\
2014 05 09	&	56786.30	&	\phs31.40	&	\ldots	&	\ldots	&	15.05 (0.03)	&	\ldots	&	\ldots	&	PS1	\\
2014 05 19	&	56796.25	&	\phs41.35	&	\ldots	&	\ldots	&	\ldots	&	14.94 (0.03)	&	\ldots	&	PS1	\\
2014 05 19	&	56796.25	&	\phs41.35	&	\ldots	&	\ldots	&	\ldots	&	14.93 (0.02)	&	\ldots	&	PS1	\\
2014 05 19	&	56796.25	&	\phs41.35	&	\ldots	&	\ldots	&	\ldots	&	14.94 (0.03)	&	\ldots	&	PS1	\\
2014 05 19	&	56796.26	&	\phs41.36	&	\ldots	&	\ldots	&	\ldots	&	14.92 (0.03)	&	\ldots	&	PS1	\\
2014 05 21	&	56798.88	&	\phs43.98	&	15.91 (0.04)	&	15.02 (0.08)	&	14.84 (0.06)	&	14.80 (0.05)	&	\ldots	&	LT	\\
2014 05 22	&	56799.24	&	\phs44.34	&	\ldots	&	\ldots	&	\ldots	&	\ldots	&	14.99 (0.06)	&	PS1	\\
2014 05 22	&	56799.24	&	\phs44.34	&	\ldots	&	\ldots	&	\ldots	&	\ldots	&	14.95 (0.03)	&	PS1	\\
2014 05 22	&	56799.24	&	\phs44.34	&	\ldots	&	\ldots	&	\ldots	&	\ldots	&	14.98 (0.07)	&	PS1	\\
2014 05 22	&	56799.24	&	\phs44.34	&	\ldots	&	\ldots	&	\ldots	&	\ldots	&	14.93 (0.01)	&	PS1	\\
2014 05 22	&	56799.26	&	\phs44.36	&	\ldots	&	\ldots	&	\ldots	&	14.93 (0.03)	&	\ldots	&	PS1	\\
2014 05 22	&	56799.26	&	\phs44.36	&	\ldots	&	\ldots	&	\ldots	&	14.93 (0.05)	&	\ldots	&	PS1	\\
2014 05 22	&	56799.26	&	\phs44.36	&	\ldots	&	\ldots	&	\ldots	&	14.87 (0.06)	&	\ldots	&	PS1	\\
2014 05 22	&	56799.26	&	\phs44.36	&	\ldots	&	\ldots	&	\ldots	&	14.87 (0.07)	&	\ldots	&	PS1	\\
2014 05 22	&	56799.27	&	\phs44.37	&	15.81 (0.07)	&	\ldots	&	\ldots	&	\ldots	&	\ldots	&	PS1	\\
2014 05 22	&	56799.28	&	\phs44.38	&	15.84 (0.04)	&	\ldots	&	\ldots	&	\ldots	&	\ldots	&	PS1	\\
2014 05 22	&	56799.30	&	\phs44.40	&	15.83 (0.09)	&	\ldots	&	\ldots	&	\ldots	&	\ldots	&	PS1	\\
2014 05 23	&	56800.24	&	\phs45.34	&	\ldots	&	\ldots	&	\ldots	&	\ldots	&	14.79 (0.11)	&	PS1	\\
2014 05 24	&	56801.25	&	\phs46.35	&	\ldots	&	\ldots	&	14.83 (0.09)	&	14.73 (0.05)	&	14.70 (0.12)	&	PS1	\\
2014 05 25	&	56802.25	&	\phs47.35	&	\ldots	&	\ldots	&	\ldots	&	14.71 (0.06)	&	14.75 (0.07)	&	PS1	\\
2014 05 26	&	56803.24	&	\phs48.34	&	\ldots	&	\ldots	&	\ldots	&	\ldots	&	14.84 (0.17)	&	PS1	\\
2014 05 27	&	56804.26	&	\phs49.36	&	\ldots	&	\ldots	&	14.69 (0.11)	&	14.69 (0.09)	&	14.87 (0.06)	&	PS1	\\
2014 05 27	&	56804.88	&	\phs49.98	&	15.86 (0.10)	&	14.98 (0.03)	&	14.86 (0.06)	&	14.80 (0.07)	&	\ldots	&	LT	\\
2014 05 30	&	56807.89	&	\phs52.99	&	15.87 (0.06)	&	14.93 (0.03)	&	14.78 (0.02)	&	14.76 (0.09)	&	\ldots	&	LT	\\
2014 05 31	&	56808.26	&	\phs53.36	&	15.95 (0.03)	&	15.05 (0.07)	&	14.73 (0.09)	&	14.77 (0.08)	&	14.83 (0.08)	&	PS1	\\
2014 06 01	&	56809.26	&	\phs54.36	&	15.99 (0.04)	&	15.11 (0.12)	&	14.83 (0.11)	&	\ldots	&	14.68 (0.10)	&	PS1	\\
2014 06 01	&	56809.92	&	\phs55.02	&	15.83 (0.08)	&	14.90 (0.06)	&	14.76 (0.03)	&	14.70 (0.05)	&	\ldots	&	LT	\\
2014 06 02	&	56810.26	&	\phs55.36	&	15.97 (0.04)	&	15.12 (0.09)	&	14.81 (0.05)	&	14.79 (0.09)	&	14.74 (0.10)	&	PS1	\\
2014 06 03	&	56811.26	&	\phs56.36	&	16.00 (0.04)	&	15.05 (0.07)	&	14.87 (0.08)	&	14.72 (0.07)	&	14.70 (0.09)	&	PS1	\\
2014 06 03	&	56811.89	&	\phs56.99	&	15.84 (0.09)	&	14.92 (0.07)	&	14.78 (0.04)	&	14.71 (0.05)	&	\ldots	&	LT	\\
2014 06 04	&	56812.26	&	\phs57.36	&	15.98 (0.03)	&	15.07 (0.09)	&	14.84 (0.05)	&	14.69 (0.06)	&	14.80 (0.07)	&	PS1	\\
2014 06 05	&	56813.89	&	\phs58.99	&	15.94 (0.07)	&	14.88 (0.06)	&	14.74 (0.03)	&	14.67 (0.08)	&	\ldots	&	LT	\\
2014 06 06	&	56814.26	&	\phs59.36	&	16.04 (0.09)	&	15.21 (0.10)	&	\ldots	&	14.80 (0.12)	&	14.75 (0.08)	&	PS1	\\
2014 06 07	&	56815.26	&	\phs60.36	&	16.00 (0.04)	&	15.06 (0.07)	&	14.78 (0.06)	&	14.69 (0.07)	&	14.73 (0.10)	&	PS1	\\
2014 06 08	&	56816.26	&	\phs61.36	&	16.06 (0.04)	&	15.13 (0.09)	&	14.87 (0.11)	&	14.76 (0.11)	&	14.77 (0.10)	&	PS1	\\
2014 06 09	&	56817.26	&	\phs62.36	&	16.06 (0.04)	&	\ldots	&	\ldots	&	14.75 (0.11)	&	\ldots	&	PS1	\\
2014 06 10	&	56818.26	&	\phs63.36	&	16.08 (0.05)	&	15.02 (0.07)	&	14.81 (0.07)	&	14.75 (0.06)	&	14.74 (0.09)	&	PS1	\\
2014 06 11	&	56819.27	&	\phs64.37	&	16.07 (0.04)	&	15.02 (0.10)	&	14.86 (0.10)	&	14.75 (0.11)	&	14.82 (0.12)	&	PS1	\\
2014 06 12	&	56820.26	&	\phs65.36	&	16.05 (0.04)	&	15.07 (0.08)	&	14.81 (0.06)	&	14.71 (0.16)	&	14.78 (0.10)	&	PS1	\\
2014 06 13	&	56821.26	&	\phs66.36	&	16.05 (0.04)	&	15.02 (0.07)	&	14.83 (0.10)	&	14.70 (0.08)	&	14.70 (0.13)	&	PS1	\\
2014 06 14	&	56822.27	&	\phs67.37	&	16.06 (0.04)	&	15.01 (0.08)	&	14.86 (0.06)	&	14.70 (0.08)	&	14.75 (0.10)	&	PS1	\\
2014 06 15	&	56823.26	&	\phs68.36	&	16.09 (0.07)	&	15.17 (0.10)	&	\ldots	&	\ldots	&	14.86 (0.15)	&	PS1	\\
2014 06 16	&	56824.26	&	\phs69.36	&	16.14 (0.07)	&	15.14 (0.08)	&	14.88 (0.06)	&	\ldots	&	14.73 (0.11)	&	PS1	\\
2014 06 16	&	56824.94	&	\phs70.04	&	15.95 (0.06)	&	14.91 (0.03)	&	14.74 (0.03)	&	14.64 (0.08)	&	\ldots	&	LT	\\
2014 06 21	&	56829.26	&	\phs74.36	&	16.17 (0.09)	&	15.09 (0.09)	&	14.88 (0.06)	&	14.73 (0.08)	&	14.78 (0.08)	&	PS1	\\
2014 06 24	&	56832.26	&	\phs77.36	&	16.16 (0.04)	&	15.11 (0.08)	&	14.86 (0.09)	&	14.74 (0.08)	&	14.74 (0.09)	&	PS1	\\
2014 06 27	&	56835.27	&	\phs80.37	&	16.20 (0.04)	&	15.14 (0.10)	&	14.86 (0.09)	&	14.68 (0.05)	&	14.71 (0.11)	&	PS1	\\
2014 06 29	&	56837.27	&	\phs82.37	&	16.18 (0.05)	&	15.02 (0.11)	&	14.78 (0.07)	&	14.75 (0.06)	&	14.69 (0.09)	&	PS1	\\
2014 07 02	&	56840.94	&	\phs86.04	&	15.97 (0.09)	&	14.94 (0.04)	&	14.72 (0.07)	&	14.62 (0.10)	&	\ldots	&	LT	\\
2014 07 06	&	56844.92	&	\phs90.02	&	16.04 (0.07)	&	14.90 (0.04)	&	14.69 (0.04)	&	14.70 (0.10)	&	\ldots	&	LT	\\
2014 07 13	&	56851.89	&	\phs96.99	&	16.20 (0.08)	&	14.96 (0.06)	&	14.77 (0.05)	&	14.77 (0.11)	&	\ldots	&	LT	\\
2014 07 18	&	56856.89	&	101.99	&	16.17 (0.10)	&	14.98 (0.06)	&	14.78 (0.08)	&	14.70 (0.13)	&	\ldots	&	LT	\\
2014 07 24	&	56862.92	&	108.02	&	16.29 (0.08)	&	15.00 (0.09)	&	14.83 (0.07)	&	14.65 (0.14)	&	\ldots	&	LT	\\
2014 08 07	&	56876.90	&	122.00	&	16.56 (0.15)	&	15.14 (0.09)	&	14.91 (0.06)	&	14.80 (0.15)	&	\ldots	&	LT	\\
2014 10 31	&	56961.26	&	206.36	&	> 19.9	&	19.23 (0.33)	&	18.17 (0.45)	&	18.05 (0.25)	&	\ldots	&	LT	\\
 \hline
\end{tabular}
\end{table}

\begin{table}
\caption{Details of recent type IIP SNe in host galaxies with Cepheid distance measurements.}
\label{table:IIP_cepheid} 
\centering
\begin{tabular}{lcccccccc}
\hline\hline 
SN 	& Galaxy &	$v_{\mathrm{hel}}^*$ & $D_{\mathrm{Cepheid}}$	&	$E(B-V)$ 	&	$m_{V_{50}}$ & $m_{I_{50}}$ &	$v_{50}^{**}$ &  References \\
& & (km s$^{-1}$) &	(Mpc) & (mag) & (mag) & (mag) & (km s$^{-1}$) &  \\
\hline
2004dj & NGC 2403 & \phs133 & \phs3.22 (0.15)  & 0.100   & 12.04 (0.03) & 11.40 (0.03) & 3250 (250) & 1, 2, 3 \\
2008bk & NGC 7793 & \phs230 & \phs3.44 (0.15)  &  0.021  & 12.85 (0.10)  & 12.10 (0.10) & 2200	(200) & 4, 5, 6 \\
2014bc & NGC 4258 & \phs448 & \phs7.40 (1.16)  & 0.014  & 15.32 (0.10) & 14.60 (0.12) & 1460 (100) & 7 \\
2012aw & NGC 3351 & \phs778 & 10.00 (0.41) & 0.086  & 13.54 (0.02) & 12.82 (0.05) & 4250 (200) & 1, 8, 9 \\
1999em & NGC 1637 & \phs717  & 11.70 (1.00)  & 0.100   & 13.98 (0.05) & 13.35 (0.05) & 3757 (300) & 10, 11, 12 \\
2001du & NGC 1365 & 1636 & 17.95 (0.41) & 0.170   & 15.00 (0.10) & 14.35 (0.10) & 3980 (220) & 1, 13, 14 \\
\hline
\end{tabular}
\tablefoot{
The SNe are ordered in terms of increasing distance.
*Heliocentric recession velocity from NED.
**Velocities extrapolated to 50 days using equation 2 of \citet{Nugent_2006}. 
1) Distance from \citet{freedman:01};
2) Magnitudes and velocity from \citet{Vinko_2006};
3) Reddening from \citet{Meikle_2011};
4) Distance from \citet{Pietrzyski_2010};
5) Magnitudes from \citet{Pignata_2013};
6) Velocity measured from spectrum downloaded from the ESO archive;
7) Distance from \citet{Fiorentino_2013};
8) Magnitudes, reddening and velocity from \citet{DallOra_2014};
9) Velocity measured from spectrum downloaded from Weizmann Interactive Supernova Data Repository (WISeREP; \citealt{Yaron_2012}; http://www.weizmann.ac.il/astrophysics/wiserep/);
10) Distance from \citet{Leonard_2003};
11) Magnitudes and velocity from \citet{Hamuy_2003};
12) Reddening from \citet{elmhamdi:03};
13) Velocity measured from spectrum in \citet{Smartt_2003} who also estimate the reddening;
14) Magnitudes from \citet{VanDyk_2003}.
}
\end{table}

\begin{table}
\caption{Shift in zero points of the 4 SCM calibrations as a result of fixing the SN distance to the respective Cepheid distance to the host galaxy.}
\label{table:IIP_ZPs} 
\centering
\begin{tabular}{lcccc}
\hline\hline 
SN 	& $\Delta\mu$ \citep{Hamuy_2003}$^\dagger$ &	$\Delta\mu$ \citep{Nugent_2006}	& $\Delta\mu$ \citep{Poznanski_2009}	& $\Delta\mu$ \citep{DAndrea_2010}  \\
& (mag) & (mag) & (mag) & (mag) \\
\hline
2004dj & \phs0.24	&	\phs0.33	&	\phs0.22	&	\phs0.43	\\
2008bk & \phs0.02	&	\phs0.05	&	$-0.05$	&	\phs0.22	\\
2014bc & $-0.10$	&	$-0.19$	&	\phs0.02	&	\phs0.38	\\
2012aw & $-0.14$	&	\phs0.07	&	$-0.37$	&	$-0.21$	\\
1999em & $-0.27$	&	$-0.17$	&	$-0.34$	&	$-0.16$	\\
2001du & $-0.19$	&	\phs0.07	&	$-0.18$	&	$-0.01$	\\
\hline
\end{tabular}
\tablefoot{The SNe are ordered in terms of increasing distance. $\Delta\mu$ is given in the sense (SCM$-$Cepheid). 
$^\dagger$\citet{Hamuy_2003} Eqn 4. 
The shifts for SN~2014bc if using the maser distance value are given in Table \ref{table:distances}.
}
\end{table}

\end{document}